\documentclass[proceedings]{JHEP} 
\usepackage{latexsym}
\font\tenmsbm=msbm10 scaled 1200
\font\sevenmsbm=msbm9
\newfam\msbmfam
\textfont\msbmfam=\tenmsbm
\scriptfont\msbmfam=\sevenmsbm
\def\msbm{\fam\msbmfam\tenmsbm}

\newcommand{\dfrac}{\displaystyle \frac}

\conference{Quantum aspects of gauge theories, supersymmetry and unification.}

\title{Three dimensional Conformal Field Theories
from Sasakian seven-manifolds\thanks{Talk presented at the TMR conference
in Paris, September 1999.}}

\author{D. Fabbri
\\
Dipartimento di Fisica Teorica, Universit\`a  di Torino,  \\
via P. Giuria 1, I-10125 Torino.}

\abstract{We present the construction of the candidate conformal field
theories dual to $AdS_4$ non-maximally supersymmetric compactifications
of 11D supergravity.
We compare the spectra of the two theories and discuss the realization
of the baryonic symmetries.
Finally we comment the presence in the spectrum of long multiplets with
rational energies, trying to explain their existence.}


\begin{document}
%
%
\section{Introduction}

One of the most exciting aspects of string theory that has emerged
over the past decade is the deep interplay between geometry and physics.
The AdS/CFT correspondence \cite{M,GKP,W,MAGOO} provides us the
possibility of testing the connection between some geometrical
features of the supergravity (SUGRA) compactifying manifold and
the field content of the dual superconformal field theory (SCFT).

The purpose of \cite{3D}, on which this talk is based, is twofold.
On one side we intend to identify the gauge theory living on
the boundary of $AdS_4$, whose IR fixed point should realize the SCFT
dual to 11D SUGRA compactified on $AdS_4\times (G/H)^7$, where $(G/H)^7$
is a homogeneous seven manifold.
On the other side, we want to check the correspondence by comparing
the spectra of the two theories.

Especially in the case of M-theory, where we have no deep control on
the fundamental dynamics, the geometrical hints are essentially the
only guidelines that help us in the construction of the worldvolume
theory of a collection of branes placed at a conifold singularity
in transverse space, namely the vertex of the metric cone over $(G/H)^7$.
We will see that when this cone, ${\cal C}(G/H)$, admits a {\it toric}
description, it is possible to argue the gauge group and the matter
field content of this theory.

We are interested in the ${\cal N}\geq 2$ supersymmetric compactifications,
which were classified in 1984 \cite{CRW}.
The possible internal manifolds are the sasakian \cite{Fig}
and trisasakian cosets:
\begin{eqnarray*}
\left.\begin{array}{cc}
M^{111}\equiv&\dfrac{SU(3)\times SU(2)\times U(1)}{SU(2)\times U(1)\times U(1)}\\
&\\
Q^{111}\equiv&\dfrac{SU(2)\times SU(2)\times U(2)}{U(1)\times U(1)}\\
&\\
V_{5,2}\,\,\equiv&\dfrac{SO(5)}{SO(3)}
\end{array}\right\}\to{\cal N}\!=\!2\\
\begin{array}{c}
N^{010}\equiv\qquad\dfrac{SU(3)}{U(1)}
\end{array}\qquad\qquad\qquad\to{\cal N}\!=\!3
\end{eqnarray*}
Contrary to the type IIB case of N $D3$-branes at a conifold
singularity \cite{KW,G}, where the existence of a nontrivial
superpotential provides a constraint on the observable
operators of the boundary SCFT, for the gauge theory of N
$M2$-branes, relevant to this work, it seems to be no such
a superpotential to help us in the identification of the
effective degrees of freedom at the strong coupling conformal point.

Despite this lack, once again geometry comes in our aid.
Indeed it is possible to recognize a common algebraic structure
between the pattern of the KK short multiplets and the description
of the transverse cone ${\cal C}(G/H)$ as an algebraic (sub)manifold.
Relying on this analogy, it is possible to associate to the generators
of the polynomial ring of ${\cal C}(G/H)$ particular combinations
of fundamental fields of the gauge theory.
These combinations seem to be the effective degrees of freedom of
the CFT, generating the {\it chiral ring} of conformal chiral
operators, at the base of the spectrum.

In this way we are able to identify the CFT boundary operators dual
to all the bulk short multiplets, matching not only the conformal
dimensions of the first with the energies of the latter, but
even the flavor and $R$-symmetry quantum numbers of each couple.
In this sense, the richer structure of the spectra of non-maximally
supersymmetric theories guarantees a less trivial test of duality
than in the case of spherical compactifications.

One further highly non-trivial check is given by the SUGRA predictions
on the anomalous conformal dimensions of the fundamental fields
of the boundary theory.
These predictions are based on the global baryonic symmetries of
the latter \cite{WHom,GK}, which reflect the existence of non-trivial
homology cycles of the internal $(G/H)^7$ \cite{DFBetti}.

Finally, an interesting aspect of these non-maximally supersymmetric
AdS/CFT pairs, seems to be the existence of some intriguing quantum
mechanism that prevents certain long multiplets of the spectrum
from acquiring anomalous dimensions.
On the base of a simple consideration on the spectra of different
supersymmetric compactifications, we will try to hint a possible
direction towards the solution of the puzzle.
%
%
\section{Identification of the gauge theory}
We want now to describe the fundamental steps in the search
for the worldvolume thoery of N $M2$-branes sitting at the vertex
of ${\cal C}(G/H)$.
We will focus on the cases where the cone is a
toric manifold, namely for $(G/H)^7=M^{111}$ or $Q^{111}$.
Hence we are looking for an ${\cal N}\!=\!2$ supersymmetric gauge
theory in three dimensions, whose moduli space of vacua should
reproduce the one of the brane system.
We make a minimality hypothesis about the matter field content:
we assume that the fundamental fields, apart from the gauge bosons,
transform in the most basic\footnote{in the sense that all the
higher representation are realizable as tensor products of this one.}
representation of the $Osp({\cal N}|4)$ superconformal group: the
{\it supersingleton}, which has a field-theoretic realization as a
particularly constrained chiral superfield.

The general structure of a three dimensional ${\cal N}\!=\!2$
supersymmetric gauge theory coupled to chiral matter is severly
constrained.
The only freedom we have is in the choice of
\begin{itemize}
\item the {\bf gauge group};
\item the scalar K\"ahler manifold, namely (in the renormalizable
flat case) the number and flavor representations of the matter
{\bf chiral multiplets};
\item the holomorphic {\bf superpotential}.
\end{itemize}
Let us begin with identifying the abelian gauge theory living on a
single brane.
Being the gauge fields frozen at the conformal point ($g_{_{Y\!M}}$
is dimensionful), we can neglect the Coulomb branch and
focus on the Higgs branch of the moduli space, parametrized by the
{\it vev} of the scalars in the chiral multiplets.
This branch should reproduce the space of vacua of the
$M2$-brane, which is the transverse space ${\cal C}(G/H)$.

This requirement is easily achieved when the cone admits a toric
description.
Roughly speaking, this means that ${\cal C}(G/H)$ can be viewed
as the K\"ahler quotient:
\begin{equation}
{\cal C}(G/H)=\dfrac{{\msbm C}^p}{({\msbm C}^*)^{(p-4)}}\label{toric}
\end{equation}
for some $p\geq 4$.
In this case, we can take as chiral fields the $p$ coordinates of
${\msbm C}^p$.
The flavor quantum numbers will be determined by the natural lifting
on ${\msbm C}^p$, of the isometries of $(G/H)^7$.
By choosing as abelian gauge group the $U(1)^{(p-4)}$ compact part of
the modding $({\msbm C}^*)^{(p-4)}$, we obtain as $D$-term equations
its non-compact ${\msbm R}^{^+\!(p-4)}$ part, in such a way to
reproduce (\ref{toric}) as the moduli space of vacua of the gauge theory.

Let us see how this construction is concretely implemented in the
case where $(G/H)^7$ is the coset $M^{111}$, whose metric cone
admits the following toric description:
\begin{equation}
{\cal C}(M^{111})=\dfrac{{\msbm C}^{^5}}{{\msbm C}^{^*}}\,.\label{M111toric}
\end{equation}
The quotient is defined by the equivalence:
\begin{equation}
(U^i,V^A)\sim(\lambda^2 U^i,\lambda^{-3}V^A)\,,\ \ 
\lambda\in{\msbm C}^{^*},\label{equivalence}
\end{equation}
where $U^i$ and $V^A$ ($i=1,2,3$, $A=1,2$) are the five
coordinates of ${\msbm C}^{^5}$, transforming respectively in
the fundamental representation of the $SU(3)$ and of the $SU(2)$
flavor factors.
The ${\msbm R}^{^+}\subset{\msbm C}^{^*}$ modding action can
be fixed by imposing:
\begin{equation}
\sum_i\vert U^i\vert^2=\sum_A\vert V^A\vert^2\,,
\end{equation}
while the $U(1)$ part of (\ref{equivalence}) determines the
charges of the corresponding chiral fields (2 and -3 respectively).

With this choice of gauge group and chiral fields, the $D$-term
of the bosonic potential is given by:
\begin{equation}
{\cal U}(z,\bar z)=\left(\sum_i\vert u^i\vert^2-\sum_A\vert v^A\vert^2
\right)^2\,,
\end{equation}
whose minimization, together with the gauge equivalence:
\begin{equation}
(u^i,v^A)\sim(e^{2i\theta}u^i,e^{-3i\theta}v^A)
\end{equation}
exactly reproduces the equation of the cone (\ref{equivalence}).
For symmetry reasons, it is worth to introduce a second $U(1)$
group, yielding the following couples of charges for the
fundamental supersingletons:
\begin{equation}
\left\{\begin{array}{cc}
U^i:&(2,-2)\\
V^A:&(-3,3)
\end{array}\right.\,.
\end{equation}
The diagonal factor will actually decouple, leaving the
gauge group $U(1)^2/U(1)_{diagonal}$.

The non-abelian generalization of this gauge theory is easily
obtained by promoting the $U(1)$ gauge groups to $SU(N)$,
with chiral matter in the following color representations:
\begin{equation}
\left\{\begin{array}{cc}
U^i:&\bf{N}^{\otimes_s 2}\otimes\bar{\bf{N}}^{\otimes_s 2}\\
V^A:&\bar{\bf{N}}^{\otimes_s 3}\otimes \bf{N}^{\otimes_s 3}
\end{array}\right.\,.
\end{equation}
%
%
\section{The conformal theory}
Let us now consider the observable fields at the IR fixed point
of this theory, where the gauge fields are integrated out.
First of all they will reduce to the gauge-invariant composite
operators, whose smallest holomorphic combination (in the abelian case)
is given by
\begin{equation}
X^{ijkAB}\equiv U^i U^j U^k V^A V^B\,,\label{X}
\end{equation}
transforming in the $(\bf{10},\bf{3})
$ of $SU(3)\times SU(2)$.

An alternative description of ${\cal C}(M^{111})$ is
in terms of an algebraic submanifold of ${\msbm C}^{30}$, where
the complex space can be parametrized by the
$10\times 3=30$ $X^{ijkAB}$ of (\ref{X}).
The embedding is given by 325 quadratic homogeneous equations in these
coordinates $X$.
In other words, on the defining locus of the cone certain quadratic
combinations of the $X^{ijkAB}$ identically vanish.
This fact is quite general and reveals a deep connection between
algebraic geometry and representation theory.
Without entering into mathematical details, it is worth to stress
that the embedding equations are equivalent to putting to zero
certain irreducible representations spanned by the quadratic
products of the $X$'s.
Indeed the symmetric tensor product:
\begin{equation}
X^{ijkAB}\cdot X^{lmnCD}\ \leftrightarrow\
(\bf{10},\bf{3})\otimes_s(\bf{10},\bf{3})\label{branch}
\end{equation}
branches into several irreducible representations of the flavor group
$SU(3)\times SU(2)$.
Among these, only the highest weight survives, while the other
constitute just the 325 combinations that, once equated to zero,
yield the embedding equations of the cone.

In the $M^{111}$ case, the only surviving
polynomials of homogeneous degree 6 and 4, respcetively in the
$U$'s and the $V$'s, are those combinations belonging
to the completely symmetric flavor representation:
\begin{equation}
\Box\!\Box\!\Box\!\Box\!\Box\!\Box_{SU(3)}\otimes
\Box\!\Box\!\Box\!\Box_{SU(2)}\,.
\end{equation}
If we consider higher degree polynomials in the $X$'s we find
that, due to the vanishing of certain quadratic subfactors, the
only surviving combinations are always those completely
symmetrized in the flavor indices.

The mathematical structure underlying these {\it selection rules}
is that of a ring.
On one side we have the polynomial ring characterizing the cone
${\cal C}(M^{111})$ as an algebraic manifold:
\begin{equation}
\dfrac{{\msbm C}[X^{ijkAB}]}{{\cal I}_{325}}\,,
\end{equation}
where ${\cal I}_{325}$ is the ideal of the embedding equations.
On the other side we have the isomorphic chiral ring of the
holomorphic color singlet operator products, modded out by the
proper ideal of vanishing relations.

This ring structure is easily extended to the non-abelian generalization
of the gauge theory.
Indeed it is possible to show that there is always only one possible
way to contract the gauge indeces into a color singlet, for each
homogeneous polynomial in the $X$'s, completely flavor symmetrized.

It is important to remark the difference between our coset
compactifications and the $T^{11}$ case of \cite{KW}.
While for the $Q^{111}$ and $M^{111}$ the non-highest weight part
of (\ref{branch}) is given by a considerable set of equations
(respectively 9 and 325), in the $T^{11}$ case the only vanishing
quadratic combination of the $X$'s corresponds to the singlet
representation of the flavor group $SU(2)\times SU(2)$ which,
in the non-abelian case, turns out to be just the superpotential
of the boundary theory.
In the other $M$-theory cases, there is no way to combine
all the non-highest weight pieces of (\ref{branch}) into a
single flavor invariant with the right dimensions for a
superpotential.
This is the fundamental reason why the vanishing relations,
which select the observable operators at the conformal point,
are to be found in a highly non-perturbative quantum mechanism,
other then the minimization of the potential.
%
%
\section{Comparison with the KK spectrum}
The first test of correspondence between the CFT and the compactified
SUGRA is the matching of the spectra of the two theories.

The KK compactifications of 11D SUGRA were extensively studied
in the first half of the 80's.
In particular, the coset space $M^{111}$ was widely investigated
\cite{WM111,CDFM111} for its seemingly possibility
to explain the origin of the $SU(3)\times SU(2)\times U(1)$
gauge bosons of the standard model.

The developement of techniques such as harmonic analysis on
coset spaces \cite{libro}, allowed the computation of great
part of the spectrum of these compactifications.
But the problems concerning the fermionic spectrum (first of all
the absence of chirality) and the occurence of the first string
revolution rapidly distorted the attention from KK SUGRA.

Due to the renewed interest, in connection to the AdS/CFT
conjecture, the KK spectra of the $M^{111}$, $Q^{111}$ and
$V_{5,2}$ compactifications have been recently completed in a
systematic way  \cite{M111,Stiefel}.

We will focus in particular on the CFT operators of protected
conformal dimensions, corresponding to KK states belonging to
short representations of the $Osp(2|4)$ symmetry superalgebra.
The following table lists all the different kinds of
$BPS$-saturated representations and the supercovariant
differential constraint implementing the corresponding
shortening conditions on the boundary superfields \cite{F,Osp}.
\[
\begin{array}{|c|c|}
\hline
\mbox{ multiplet} & \mbox{ differential~constraint~on} \\
&\mbox{ the~boundary~superfield}\\
\hline
{\rm short~graviton} & {\cal D}^{+\alpha}
\Phi_{(\alpha\beta)}\left(x,\theta^\pm\right)=0\\
\hline
{\rm short~gravitino} & {\cal D}^{+\alpha}
\Phi_{\alpha}\left(x,\theta^\pm\right)=0\\
\hline
{\rm short~vector} & {\cal D}^{+\alpha}{\cal D}^+_{\alpha}
\Phi\left(x,\theta^\pm\right)=0\\
\hline
{\rm hypermultiplet} & {\cal D}^{+\alpha}
\Phi\left(x,\theta^\pm\right)=0\\
\hline
{\rm massless~graviton} & \left\{\begin{array}{c}
{\cal D}^{+\alpha}
\Phi_{(\alpha\beta)}\left(x,\theta^\pm\right)=0\\
{\cal D}^{-\alpha}
\Phi_{(\alpha\beta)}\left(x,\theta^\pm\right)=0
\end{array}\right.\\
\hline
{\rm massless~gravitino}  & \left\{\begin{array}{c}
{\cal D}^{+\alpha}
\Phi_{\alpha}\left(x,\theta^\pm\right)=0\\
{\cal D}^{-\alpha}\Phi_{\alpha}\left(x,\theta^\pm\right)=0
\end{array}\right.\\
\hline
{\rm massless~vector} & \left\{\begin{array}{c}
{\cal D}^{+\alpha}{\cal D}^+_{\alpha}
\Phi\left(x,\theta^\pm\right)=0\\
{\cal D}^{-\alpha}{\cal D}^-_{\alpha}
\Phi\left(x,\theta^\pm\right)=0
\end{array}\right.\\
\hline
{\rm supersingleton} & \left\{\begin{array}{c}
{\cal D}^{+\alpha}
\Phi\left(x,\theta^\pm\right)=0\\
{\cal D}^{-\alpha}{\cal D}^-_{\alpha}
\Phi\left(x,\theta^\pm\right)=0
\end{array}\right.\\
\hline
\end{array}
\]
For space reasons, it is not possible here, to make a complete
list of the protected conformal superfields corresponding to all
the short multiplets of the KK spectrum.
We will focus on the most meaningful.

As we anticipated, the chiral ring of the SCFT
is given by the following composite operators:
\begin{equation}
{\it Tr}\left[(U^3V^2)^k\right]\,,\label{chiral}
\end{equation}
where, following the prescription given in the last section,
the flavor indices are completely symmetrized, while the trace
symbol implies the only possible contraction into a color-singlet,
compatible with this flavor representation.
These chiral superfields find a complete matching with the
tower of hypermultiplets of the KK spectrum.
The only constraint we have, is on the anomalous dimensions of the
fundamental supersingletons, which we are not able to directly deduce
from the CFT:
\begin{equation}
2=\Delta[Tr(U^3V^2)]=3\Delta(U)+2\Delta(V)\,.\label{Delta}
\end{equation}
In the next section we will show how $\Delta(U)$ and $\Delta(V)$
can be obtained from SUGRA computations, confirming the validity
of our conjecture on the dual CFT.

Other important superfields are the conserved supercurrents:
\begin{equation}
^{SU(3)}J^i_{\ j}=Tr\left[U^i\overline U_j
-\dfrac{1}{3}\delta^i_{\ j}U^k\overline U_k\right]\label{SU(3)J}
\end{equation}
and
\begin{equation}
^{SU(2)}J^A_{\ B}=Tr\left[\overline V^AV_{\ B}
-\dfrac{1}{2}\delta^A_{\ B}\overline V^CV_C\right]\,,\label{SU(2)J}
\end{equation}
that transform in the adjoint representation of the flavor groups,
$SU(3)$ and $SU(2)$ respectively, and contain the massless vectors
associated to these symmetries.
Two sets of semiconserved currents, corresponding to short
vector multiplets of the KK spectrum are given by composing (\ref{SU(3)J})
and (\ref{SU(2)J}) with the chiral fields (\ref{chiral}).

Finally, two towers of KK short gravitinos are matched by the following
spinorial superfields:
\[
Tr\left[\left(U\overline{U}({\cal D}^+_\alpha\overline{V}V)
+\overline{V}V({\cal D}^+_\alpha U\overline{U})\right)^{i\ A}_{\ j\ B}
(U^3V^2)^k\right]
\]
and
\[
Tr\left[\left(U^iU^jU^kV^A{\cal D}^-_\alpha V^B\epsilon_{AB}\right)
(U^3V^2)^k\right]\,.
\]
All these superfields and the other which we have not listed, perfectly
fit the masses, $R$-symmetry charges and flavor quantum numbers of
the corresponding KK short multiplets.
%
%
\section{The baryonic symmetries}
The next item we want to discuss is the presence in the boundary
theory of barion-like conformal operators and their interpretetion in
terms of non-perturbative bulk states.

Already in the gold years of KK SUGRA, it was noted \cite{DFBetti}
that the existence of non trivial homological cycles in the
internal manyfold implies the presence of massless
vectors in the spectrum of the compactified SUGRA,
due to the reduction of higher degree form potential on these cycles.
The corresponding {\it Betti multiplets} have been related, in the
AdS/CFT perspective, to global baryonic symmetries of the dual
gauge theory \cite{WHom}.

In the specific case considered in this talk, it is found that the
manyfold $M^{111}$ has non vanishing second and fifth Betti numbers:
$b_2(M^{111})=b_5(M^{111})=1$, implying a $U(1)$ baryonic symmetry.
All the KK multiplets (and hence all the dual conformal operators)
result discharged under this symmetry.
But we have not analysed, so far, all the possible color singlet
products of fundamental supersingletons at our disposal.
Among these, we have the totally antisymmetrized\footnote{implying
total symmetrization in the flavor indices}
(in the $SU(N)$ color indices) product of $N$ singletons of the
same kind, that we will abbreviate $det(U)$ and $det(V)$.
The masses of these operators grow like $N$ in the large $N$ limit.
Hence their SUGRA duals have to be found among non-perturbative
states.
Indeed we find the existence in $M^{111}$ of two particular
families of non trivial supersymmetric 5-cycles, over which
a solitonic $M5$-brane can be wrapped.
The energy of these $M$-theory configurations, simply related
to the volume of the cycles, has the same large $N$ behaviour.
The exact coefficient in this linear expression can be used as
a SUGRA prediction on the conformal anomalous dimension of
the corresponding fundamental singleton.
In this case we find:
\begin{eqnarray*}
\Delta[det(U)]=\dfrac{4N}{9}\Longrightarrow\Delta(U)=\dfrac{4}{9}\\
\Delta[det(V)]=\dfrac{N}{3}\Longrightarrow\Delta(V)=\dfrac{1}{3}
\end{eqnarray*}
These numbers perfectly fit the only constraint (\ref{Delta}) we
have from the SCFT side:
\begin{equation}
3\Delta(U)+2\Delta(V)=2\,.
\end{equation}
Furthermore, even the flavor irreducible representation of
these configurations (deducible from the action of the $M^{111}$
isometries on the cycles of each family) perfectly agree with the
flavor quantum numbers of the corresponding {\it determinant}
operators, yielding a highly non trivial check of duality. 
%
%
\section{Comments about the rational long multiplets}
An interesting feature common to both type IIB and 11D SUGRA coset
compactifications, first noted in \cite{T11} for the $T^{11}$ space,
is the presence in the KK spectrum of long multiplets with rational
energies (see also the talk by G. Dall'Agata about this point).
This fact does not seem to be a mere coincidence.
Indeed it has been recognized \cite{T11} that the dual conformal
operators follow a precise pattern: they are products of two or
more quantities separately protected that, nevertheless, have no
{\it a priori} reason to be globally protected.
The conformal dimension of these operator products, as it is deduced
by the energy of the corresponding SUGRA states, is instead given
by the {\it naive} sum of the dimensions of the single factors.

To make a concrete example, both the $M^{111}$ and $Q^{111}$ bulk
spectra contain a tower of long graviton multiplets, associated to
boundary vector superfields of the form:
\begin{equation}
\Phi_{(\alpha\beta)}\sim T_{(\alpha\beta)}\times J\times\phi
\,,\label{rational}
\end{equation}
where $T_{(\alpha\beta)}$, $J$ and $\phi$ are respectively
the stress energy tensor, a conserved vector current and a chiral
operator, and the dimension of $\Phi$ is precisely given by:
\begin{equation}
\Delta(\Phi)=\Delta(T)+\Delta(J)+\Delta(\phi)\,.
\end{equation}

We have not yet a definitive explaination of the quantum mechanism
that seems to protect these operators from acquiring anomalous
dimensions.
But a simple consideration has emerged while studying the ${\cal N}\!=\!3$
$SCFT$ corresponding to the SUGRA compactification on the
trisasakian manifold $N^{010}$ \cite{N010}.

From the analysis of the $Osp(3|4)$ supermultiplets and of their
decomposition into ${\cal N}\!=\!2$ irreducible representations \cite{N=3},
one can realize that the short ${\cal N}\!=\!3$ multiplets always contain
long multiplets of the lower superalgebra.
Furthermore, from a careful identification of the corresponding
boundary conformal operators, we have recognized for some of these
the same structure of (\ref{rational}).
Obviously, belonging to the same supermultiplet of the higher
supersymmetry algebra, these long ${\cal N}\!=\!2$ multiplets have
energies that differ from those of the other states only by
(half)-integers.
Hence they are necessarily rational, despite the fact that, from
the ${\cal N}\!=\!2$ viewpoint, they are long.

This simple consideration seems to hint that the quantum protection
mechanism previously advocated may be found in a residual form of
higher supersymmetry that the SCFT could reflect, such as a
spontaneous supersymmetry breaking.
%
%
\paragraph{Acknowledgements.}
I would like to thank all the authors of the paper \cite{3D},
on which this talk is based and also P. Termonia for his collaboration
in the preliminary results on harmonic analysis.
This work is supported by the European Commission TMR programme
ERBFMRX-CT96-0045.
%
%

\end{document}